\documentclass[conference]{IEEEtran}
\IEEEoverridecommandlockouts
\usepackage{cite}
\usepackage{amsmath,amssymb,amsfonts}
\usepackage{graphicx}
\usepackage{textcomp}
\usepackage{xcolor}
\usepackage{subcaption}
\usepackage{algorithm}
\usepackage{algpseudocode}
\usepackage{url}
\usepackage{balance}
\usepackage{soul}
\usepackage{tcolorbox}
\tcbuselibrary{breakable}
\def\BibTeX{{\rm B\kern-.05em{\sc i\kern-.025em b}\kern-.08em
    T\kern-.1667em\lower.7ex\hbox{E}\kern-.125emX}}

\begin{document}

\title{
Evaluation and Explainability of Unsupervised Scholarly Collaboration Recommendations
%via Topic, Network, and Embedding Models
}

\author{\IEEEauthorblockN{Md Asaduzzaman Noor}
\IEEEauthorblockA{Gianforte School of Computing \\
Montana State University\\
Bozeman, Montana, USA \\
mdasaduzzamannoor@montana.edu}
\and
\IEEEauthorblockN{John W. Sheppard}
\IEEEauthorblockA{Gianforte School of Computing \\
Montana State University\\
Bozeman, Montana, USA \\
john.sheppard@montana.edu}
\and
\IEEEauthorblockN{Jason A. Clark}
\IEEEauthorblockA{Library \\
Montana State University\\
Bozeman, Montana, USA \\
jaclark@montana.edu}
}

%\author{\IEEEauthorblockN{
%1\textsuperscript{st} 
%Anonymous}
%\IEEEauthorblockA{\textit{Department} \\
%\textit{Institution}\\

\maketitle

\begin{abstract}
In this paper, we examine unsupervised, content-based collaboration recommendations using publication text in scholarly settings. We compare three families of methods: a TF-IDF baseline, topic-based models (LDA and BERTopic, including clone variants), and embedding-based retrieval using SciBERT with Faiss. To evaluate model behavior beyond simple lexical matching, we introduce a constrained setting where publication overlap between researchers is partially removed while still using historical co-authorship as proxy ground truth for post-hoc evaluation.
Results show clear differences across methods. TF-IDF performs best under full information but drops significantly as overlap is reduced. In contrast, topic-based and embedding-based approaches show more stable performance, suggesting they capture broader distributional similarities, rather than relying only on direct lexical overlap. We also examine explainability through two perspectives: intrinsic topic-based explanations and post-hoc, retrieval-based explanations generated using language models. These provide complementary trade-offs between transparency and human readability.
\end{abstract}


\section{Introduction}
A perennial challenge in scholarly settings is helping scholars identify potential research collaborators.
In this work, we study unsupervised scholarly collaboration recommendation based solely on publication text (titles and abstracts). 
The goal is to identify potential research collaborations using only content-based signals, without relying on prior graph supervision during model construction. 
Collaboration recommendation is particularly challenging in interdisciplinary settings, where historical co-authorship is sparse or absent, despite strong topical similarity.

We evaluate three families of approaches. First, we apply a TF-IDF-based baseline as a lexical similarity method without any graph structure. 
% Second, we use topic model-based approaches (namely, LDA and BERTopic), along with their cloning variants, which are designed to better capture secondary research themes and mitigate publication imbalance across researchers.
Second, we use topic model-based approaches (namely, LDA and BERTopic), along with their cloning variants introduced in our prior work \cite{noorFIXscholarnodes,noorFIXhandling}, for improved researcher representation by capturing secondary research themes and mitigating publication imbalance across researchers.
Third, we include an embedding-based retrieval method using SciBERT representations with Faiss for nearest-neighbor search over researcher-level embeddings.

Evaluation of unsupervised, content-based collaboration methods remains limited. While topic-based approaches provide interpretable recommendations, their evaluation often focuses on recovering observed collaborations, with less attention to how these methods behave under reduced or incomplete publication information.

A key aspect of our study is the evaluation setting. We use historical co-authorship links as a proxy ground truth for post-hoc evaluation, while ensuring that all models remain fully unsupervised during training. To better understand how models behave with limited information, we further evaluate performance under reduced publication overlap.

For explainability, we consider two complementary approaches. Topic-based methods offer intrinsic interpretability through topic distributions, shared concepts, and word cloud-based visualizations. In contrast, for embedding-based methods, we introduce a post-hoc explanation framework that uses retrieved publication contexts and large language model-based summarization to generate human-readable rationales.

\section{Related Work} \label{sec:related}
Scholarly collaboration recommendation has traditionally relied on co-authorship networks, where existing relationships are used to predict future collaborations \cite{newman2001structure,barabasi1999emergence}. While effective at recovering known partnerships, these methods tend to reinforce existing structures and are limited in finding new or interdisciplinary collaborations.
To address this, content-based and hybrid approaches use textual and topical similarity between researchers. Topic models and similarity methods can identify related researchers even when no prior collaboration exists \cite{noorFIXidentifying,noorFIXscholarnodes,rosen2012author,yang2015evaluating,purwitasari2020identifying,hui2020hybrid}. 
More recent work uses embedding-based representations, including neural language models, to capture deeper contextual relations between publications, and their performance is often influenced by the availability of overlapping textual signals \cite{reimers2019sentence, beltagy-etal-2019-scibert}.

Despite these advances, evaluation in collaboration recommendation still relies heavily on historical co-authorship. Common metrics such as precision and recall \cite{li2024scientific,makarov2019predicting} mainly reward methods that recover existing links. 
This can undervalue methods that find relevant but previously unseen collaborations. In practice, researchers with similar work may never have collaborated, especially across institutions or fields.
Recent studies highlight the need for diversity and interdisciplinarity in recommendations \cite{yang2025enhancing,moirano2020creative}. These works suggest that evaluation should not only focus on finding known links.

Explainability is also important for recommending non-obvious collaborators. Topic-based models are naturally interpretable because they show shared topics and word distributions. Embedding-based methods are less interpretable and usually need post-hoc explanations based on retrieved documents or language models \cite{guesmi2023justification,chatti2024visualization}.

% While prior work commonly evaluates collaboration recommendation using held-out co-authorship links, author representations are typically constructed using the full set of available publications. In contrast, we restrict the publication information used to build these representations and evaluate how different methods behave under reduced textual overlap, while keeping the same evaluation links.
While prior work, including our own, has focused primarily on developing researcher representations, comprehensive evaluation under reduced publication information and comparative analysis of explainability have received less attention. In contrast, we restrict the publication information used to construct author representations and systematically evaluate lexical, topic-based, and embedding-based methods under reduced textual overlap while keeping the same evaluation links.

\section{Dataset} \label{sec:dataset}
We construct a scholarly dataset by integrating faculty rosters with publication metadata from three peer public institutions: 
% Uni1, Uni2, and Uni3.\footnote{The names have been made generic for double-blind purposes.} 
Montana State University (MSU), Washington State University (WSU), and Colorado State University (CSU).
These institutions provide comparable academic environments across multiple disciplines.
The dataset contains over 1,600 researchers and 45,000 publications spanning January 2004 to May 2025. For each publication, we use only the title and abstract as textual input for downstream modeling.

%%% Uni1 = MSU; Uni2 = WSU; Uni3 = CSU; UniC = MWC

Faculty rosters are collected from official university sources, while publication metadata is obtained from OpenAlex \cite{priem2022openalex}. Publications are matched to faculty using name-based matching with institutional constraints. 
% We also construct a combined Uni1--Uni2--Uni3 (UniC) dataset for cross-institutional analysis.
We also construct a combined MSU--WSU--CSU (MWC) dataset for cross-institutional analysis.
Standard preprocessing is applied to publication text, including lowercasing, tokenization, lemmatization, and stopword removal using NLTK and spaCy \cite{nltk,spacy}. Publications with missing or extremely short abstracts are removed, and researchers with fewer than five publications are excluded. Table~\ref{tab:dataset_summary_stats} reports summary statistics of publication counts per researcher for each institution.

\begin{table}[t!]
\centering
\caption{Dataset statistics across institutions (\#Res: number of researchers, \#Pub: number of publications)}
\label{tab:dataset_summary_stats}
\begin{tabular}{|l|c|c|c|c|c|c|}
\hline
\textbf{Institution} & \textbf{\#Res} & \textbf{\#Pub} & \textbf{Mean} & \textbf{Med} & \textbf{Max} & \textbf{Std} \\
\hline
MSU & 276 & 6{,}372  & 27.86 & 18 & 168 & 27.13 \\
WSU & 613 & 15{,}931 & 31.05 & 18 & 237 & 34.53 \\
CSU & 745 & 23{,}072 & 37.75 & 25 & 246 & 39.42 \\
MWC & 1{,}634 & 45{,}198 & 33.57 & 21 & 246 & 35.98 \\
\hline
\end{tabular}
\end{table}

\section{Methodology} \label{sec:methodology}

\subsection{Problem Formulation}

The problem we address in this work is the following: given a set of research publication metadata, specifically titles and abstracts, how can we recommend potential collaborators using only content-based signals, without relying on prior collaboration history, while promoting novel recommendations based on shared topical interests? Additionally, we evaluate the effectiveness of such recommendations in the absence of explicit ground truth.
To address these questions, we explore several unsupervised recommendation models, evaluate their effectiveness under a unified framework, and analyze how explanations can be generated from each model based on their underlying representations. 
%The different models and their corresponding representations are described below.

\subsection{Representation \& Recommendation Models}

\subsubsection{TF-IDF Baseline}

One of the earliest approaches for text-based recommendation is based on Term Frequency–Inverse Document Frequency (TF-IDF) \cite{ramos2003using}. In this approach, each document
%, in our case a publication (title + abstract), 
is represented as a numerical vector over a global vocabulary extracted from the corpus. Each dimension corresponds to a word, weighted by its TF-IDF score, where inverse document frequency down-weights commonly occurring terms and highlights document-specific keywords.

We first convert all publications into TF-IDF vectors. Since our goal is researcher-level recommendation, we aggregate and normalize the publication vectors for each researcher to obtain a researcher profile representation. Given a query researcher, we compute cosine similarity between their profile and all other researcher profiles, and return the top-$k$ most similar researchers as recommendations.
This model serves as a baseline for comparison. However, its explainability is limited, as it relies solely on word-level importance without capturing higher-level semantic structure.

\subsubsection{Topic-Based Graph Models}

We next consider topic-based recommendation models, where publications are represented using topic probability distributions. These models are based on our prior work \cite{noorFIXscholarnodes, noorFIXhandling}, which utilizes both Latent Dirichlet Allocation (LDA) and BERTopic for topic modeling.

LDA assumes that documents are mixtures of latent topics, and each topic is a distribution over words. We train LDA on the publication corpus to obtain document-level topic distributions.
BERTopic, on the other hand, follows a different pipeline. It first uses a pretrained sentence embedding model to map documents into a high-dimensional semantic space. These embeddings are then reduced using UMAP, followed by clustering via HDBSCAN to identify groups of semantically similar documents, which correspond to topics. BERTopic also provides document-topic and topic-word representations to maintain interpretability. In this work, we use a SciBERT-based sentence encoder \cite{beltagy-etal-2019-scibert} to capture scientific terminology.

To construct researcher representations, we follow different strategies for LDA and BERTopic. For LDA, we concatenate all publications of a researcher into a single document and infer a topic distribution. For BERTopic, we aggregate and normalize the topic distributions of individual publications to obtain a researcher-level representation.
Using these representations, we construct a topic similarity graph where nodes correspond to researchers. Edge weights are defined using Jensen–Shannon divergence (JSD) between topic distributions, converted to similarity as $1 - \text{JSD}$. To avoid fully connected graphs, we apply a threshold to retain only strong similarities.
We then apply nested hierarchical Louvain community detection \cite{noorFIXidentifying} to identify multi-level topical communities. Finally, we generate top-$k$ recommendations using Personalized PageRank \cite{page1999pagerank} over the constructed graph.
These models provide a structured view of researcher relationships and support more interpretable recommendation mechanisms.

\subsubsection{Clone-Based Representations}

To address limitations in representing prolific researchers, prior work \cite{noorFIXhandling} introduces a cloning strategy that captures multiple topical aspects of a researcher’s work.
Prolific researchers refer to authors with a sufficiently large publication record to exhibit multiple distinct research themes.
The selection criteria and threshold are described in detail in our previous work \cite{noorFIXhandling} and are based on distributional properties in their respective institutions.
Highly prolific researchers often have diverse research themes, but standard aggregation tends to emphasize dominant topics while suppressing secondary ones. To address this, we create multiple “clone” representations for such researchers.

For LDA-based cloning, we train a local LDA model on the publications of a prolific researcher and cluster the resulting topic distributions using HDBSCAN. Each cluster represents a distinct topical theme, and each is treated as a separate node (e.g., $A$ becomes $A_1, A_2, A_3$).
For BERTopic-based cloning, clusters are formed directly using dense publication embeddings for that researcher.

Clone nodes are incorporated into graph construction and merged back after community detection by retaining the most relevant edges. The remaining pipeline follows the topic-based graph model.

\subsubsection{Embedding-Based Retrieval}

%In addition to topic-based representations, 
Finally, we consider a dense embedding-based retrieval approach, using a pretrained SciBERT encoder to obtain vector representations for each publication.
To construct researcher-level representations, we aggregate and normalize the embeddings of all publications associated with each researcher. Given these embeddings, we perform nearest-neighbor search to retrieve top-$k$ candidate collaborators based on cosine similarity.

To enable efficient large-scale retrieval, we use Faiss \cite{johnson2019billion} for indexing and similarity search. We use a flat index with inner product search after normalizing embeddings, which makes it equivalent to cosine similarity.
This approach avoids explicit topic modeling by operating directly in a dense embedding space, while still capturing latent similarities between publications.
In addition, the embedding space is reused in the explainability framework, where retrieved publications are used as contextual evidence for generating explanations.

\subsection{Explainability Framework}

While the proposed models generate collaboration recommendations from different representations, understanding why a recommendation is made is equally important. In this work, we design explanation strategies aligned with each model’s representation, allowing recommendations to be interpreted through either structured topical similarity, relational structure, or semantic evidence.
%In general, explanations are either derived directly from the model structure, such as topic distributions or graph relationships, or constructed from retrieved publication evidence using semantic similarity.

\subsubsection{Topic-Based Explanation}

For topic-based models, explanations are derived from the topic probability distributions representing each researcher. Given a recommended researcher pair, we identify the shared topics contributing most to their similarity. For representations with multiple profiles per researcher (e.g., cloning), similarity is evaluated across all profile pairs, and the most relevant pair is selected, consistent with the recommendation process.

We compute the joint contribution of each shared topic by combining the topic probabilities of both researchers. The highest-contributing topics, together with their topic--word distributions, provide interpretable descriptions of the shared research themes underlying the recommendation.

\subsubsection{Embedding-Based Explanation}

For embedding-based recommendations, explanations are generated for individual researcher pairs. Given a query researcher $A$ and a recommended researcher $B$, we construct explanations using cross-representation similarity in the embedding space. First, aggregated and normalized publication embeddings are computed to obtain researcher-level representations.

We then retrieve two complementary sets of publications:
\textbf{(1) Query-side evidence:} For researcher $A$, we retrieve the top-$k$ publications most similar to researcher $B$'s embedding centroid, highlighting the aspects of $A$ most relevant to the recommended collaborator.
\textbf{(2) Candidate-side evidence:} For researcher $B$, we retrieve the top-$k$ publications most similar to researcher $A$'s embedding centroid, capturing the subset of $B$'s work that aligns most closely with the query researcher.

This symmetric cross-retrieval ensures that both publication sets are selected based on mutual relevance rather than isolated topical importance. The retrieved publications are then provided as contextual evidence to a large language model, which generates a natural language explanation of the shared research themes. The LLM is prompted to generate explanations exclusively from the retrieved publications, treating it as a grounded summarization step, thus constraining the explanations to their target contexts.

\subsection{Evaluation}

\subsubsection{Ground Truth Setup}

The dataset includes both publication content and co-authorship metadata from OpenAlex, allowing us to train models using only textual information while reserving co-authorship data for evaluation.
A straightforward approach would be to remove all shared publications between co-authors and test whether models can rediscover them. However, this complete removal strategy is problematic, particularly for topic-based graph models that rely on global structure. Removing all shared works can eliminate low-publication authors from the network entirely, biasing evaluation toward prolific researchers.

To address this, we adopt a partial holdout strategy. For each co-author pair $(a,b)$ with multiple shared publications, we randomly split their joint works into two halves, assigning one half to $a$ and the other to $b$. This reduces direct overlap while preserving sufficient textual signal for both authors. As a result, all researchers remain in the dataset, enabling evaluation across the full population.
We consider author pairs with at least two shared publications to avoid noise from incidental collaborations.

We report two evaluation scenarios:
% \begin{enumerate}
%     \item \textbf{Full-data evaluation:} all publications and co-authorships are retained, providing an upper-bound performance estimate.
%     \item \textbf{Holdout evaluation:} the 50/50 split is applied, testing the ability of models to recover partially hidden collaborators.
% \end{enumerate}
\textbf{(1) Full-data evaluation:} all publications and co-authorships are retained, providing an upper-bound performance estimate.
\textbf{(2) Holdout evaluation:} the 50/50 split is applied, testing the ability of models to recover partially hidden collaborators.

While past co-authorship provides a practical proxy for evaluation, it does not fully capture the space of potential collaborations. 
Researchers working on highly similar topics may not have collaborated due to institutional, temporal, or social factors. 
Thus, the absence of a co-authorship link should not be seen as a lack of relevance between two researchers.
This limitation is particularly important in our setting, where we seek to promote novel and potentially cross-disciplinary collaborations. 
As such, the reported metrics should be interpreted as measuring the ability to recover known collaborations, rather than the full set of plausible collaborations.

\subsubsection{Evaluation Metrics}

We adopt two standard ranking metrics: \textbf{Hits@k} and \textbf{Mean Reciprocal Rank (MRR)} \cite{herlocker2004evaluating}.
Hits@k measures the fraction of true collaborators retrieved within the top-$k$ recommendations:
\[
\text{Hits@k} = \frac{1}{|Q|} \sum_{q \in Q} \frac{|A_q \cap R_q^{(k)}|}{|A_q|}
\]
where $Q$ is the set of query authors, $A_q$ is the set of ground-truth collaborators, and $R_q^{(k)}$ denotes the top-$k$ recommendations for author $q$.
MRR captures the ranking position of the first relevant collaborator:
\[
\text{MRR} = \frac{1}{|Q|}\sum_{q\in Q} \frac{1}{r_q},
\]
where $r_q$ is the rank of the first true collaborator for author $q$, or zero if none are retrieved.
Hits@k reflects how often correct collaborators are retrieved within the top-$k$, while MRR emphasizes how highly they are ranked. Together, they provide complementary views of recommendation performance.

\subsubsection{Implementation Details}

For LDA, we used the Gensim implementation with the MALLET backend. The number of topics was selected using the $C_V$ coherence score, resulting in 80 (MSU), 80 (WSU), 90 (CSU), and 160 (MWC) topics for the respective datasets.

For BERTopic, we used SciBERT as the base encoder (\texttt{allenai/scibert\_scivocab\_uncased}). The model was fine-tuned using masked language modeling, and the checkpoint with the highest topic coherence was used for downstream experiments. BERTopic uses UMAP and HDBSCAN with tuning to obtain stable topic granularity across datasets.

For cloning-based representations, local topic models were constructed for selected prolific researchers, with publications grouped using density-based clustering. Researcher similarity was computed using JS divergence over topic distributions, followed by edge pruning. Community detection was performed using Nested Hierarchical Louvain with a minimum community size constraint.

For embedding-based retrieval, SciBERT-derived publication embeddings were aggregated into researcher representations using normalized averaging. Similarity search was performed using Faiss with a flat inner-product index over normalized vectors, equivalent to cosine similarity.

For explanation generation, the top-5 most relevant publications per researcher pair were retrieved using embedding similarity and provided as context to GPT-OSS-120B to generate natural language explanations.

\section{Results \& Discussion} \label{sec:results}

\begin{table*}[t!]
\centering
\caption{Full-data evaluation across institutions (MRR and Hits@10)}
\label{tab:full_results}
\begin{tabular}{l|cc|cc|cc|cc}
\hline
 & \multicolumn{2}{c|}{MSU} & \multicolumn{2}{c|}{WSU} & \multicolumn{2}{c|}{CSU} & \multicolumn{2}{c}{MWC} \\
Model & MRR & H@10 & MRR & H@10 & MRR & H@10 & MRR & H@10 \\
\hline
TF-IDF     & 0.597 & 0.915 & 0.529 & 0.849 & 0.483 & 0.809 & 0.459 & 0.745 \\
LDA        & 0.430 & 0.733 & 0.354 & 0.681 & 0.381 & 0.633 & 0.360 & 0.588 \\
Clone-LDA  & 0.416 & 0.762 & 0.385 & 0.696 & 0.391 & 0.655 & 0.373 & 0.616 \\
BERT       & 0.107 & 0.140 & 0.278 & 0.574 & 0.283 & 0.517 & 0.230 & 0.408 \\
Clone-BERT & 0.212 & 0.354 & 0.309 & 0.575 & 0.289 & 0.520 & 0.220 & 0.396 \\
Faiss      & 0.448 & 0.774 & 0.432 & 0.723 & 0.413 & 0.679 & 0.366 & 0.589 \\
\hline
\end{tabular}
\end{table*}

\begin{table*}[t!]
\centering
\caption{Holdout evaluation across institutions (MRR and Hits@10)}
\label{tab:holdout_results}
\begin{tabular}{l|cc|cc|cc|cc}
\hline
 & \multicolumn{2}{c|}{MSU} & \multicolumn{2}{c|}{WSU} & \multicolumn{2}{c|}{CSU} & \multicolumn{2}{c}{MWC} \\
Model & MRR & H@10 & MRR & H@10 & MRR & H@10 & MRR & H@10 \\
\hline
TF-IDF     & 0.408 & 0.715 & 0.397 & 0.700 & 0.366 & 0.659 & 0.207 & 0.361 \\
LDA        & 0.336 & 0.605 & 0.274 & 0.556 & 0.302 & 0.546 & 0.260 & 0.456 \\
Clone-LDA  & 0.329 & 0.629 & 0.309 & 0.603 & 0.312 & 0.563 & 0.267 & 0.471 \\
BERT       & 0.091 & 0.130 & 0.189 & 0.447 & 0.237 & 0.438 & 0.166 & 0.314 \\
Clone-BERT & 0.203 & 0.323 & 0.227 & 0.460 & 0.239 & 0.442 & 0.168 & 0.304 \\
Faiss      & 0.321 & 0.611 & 0.331 & 0.627 & 0.341 & 0.593 & 0.272 & 0.475 \\
\hline
\end{tabular}
\end{table*}

Table~\ref{tab:full_results} shows the full-data evaluation across all institutions and models.
We also evaluated Hits@k for $k \in {3, 5, 10, 20}$, observing consistent relative trends across settings; we report k = 10 for brevity.
In this setting, TF-IDF achieves the strongest performance, consistently outperforming other approaches across both MRR and Hits@10. 
This is expected, as co-authored publications share strong lexical overlap, which directly benefits term-based similarity. 
Topic-based and embedding-based methods follow, with Faiss and Clone-LDA showing the most competitive among non-lexical approaches.

\begin{figure}[t!]
    \centering
    \includegraphics[width=0.9\linewidth]{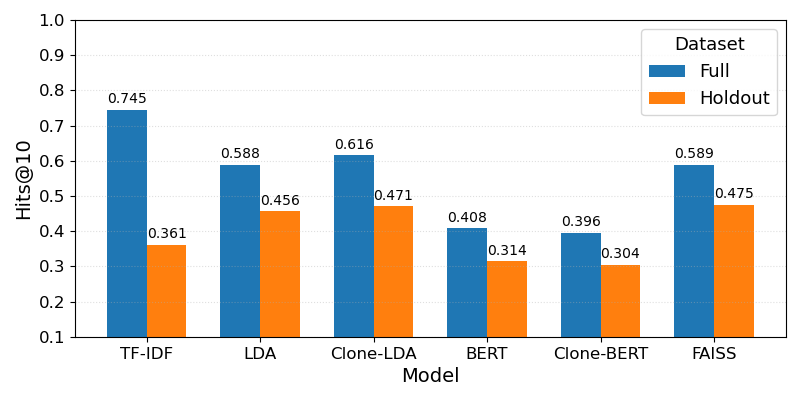}
    \caption{Hits@10 Comparison: Full vs Holdout MWC dataset}
    \label{fig:hits10_comparison}
\end{figure}

Table~\ref{tab:holdout_results} shows results under the 50/50 holdout setting, where shared co-authorship signals are partially removed. Under this setting, a shift in model behavior is observed. TF-IDF shows a notable performance degradation across all institutions, most prominently on the combined MWC dataset.
In contrast, Faiss and Clone-LDA exhibit smaller relative drops in performance, indicating greater stability under reduced lexical and co-authorship overlap. In some cases, particularly on the MWC dataset, these methods become competitive with or exceed TF-IDF in ranking performance.
Figure~\ref{fig:hits10_comparison} further illustrates this trend using Hits@10 scores. TF-IDF shows the largest degradation between full and holdout settings, whereas Faiss and Clone-LDA maintain more consistent performance.

Consistently, BERT-based variants underperformed compared to both topic-based and TF-IDF approaches. 
This may be due to BERTopic's clustering process, which tends to produce finer topic granularity than LDA, which then tends to cause BERTopic to miss broader research themes that might be in common between researchers.
However, cloning improved stability for both LDA and embedding-based representations by capturing intra-author topical variation.

Overall, the results highlight a distinction between lexical matching and representation-based approaches. While TF-IDF performs strongly when direct textual overlap is present, its effectiveness decreases under reduced overlap conditions. In contrast, topic-based and embedding-based methods show more stable performance.
%, suggesting improved robustness to distributional shift. 
%This supports the use of holdout evaluation as a more realistic setting for assessing collaboration recommendation systems.

\begin{figure*}[t!]
    \centering
    \includegraphics[width=0.6\linewidth]{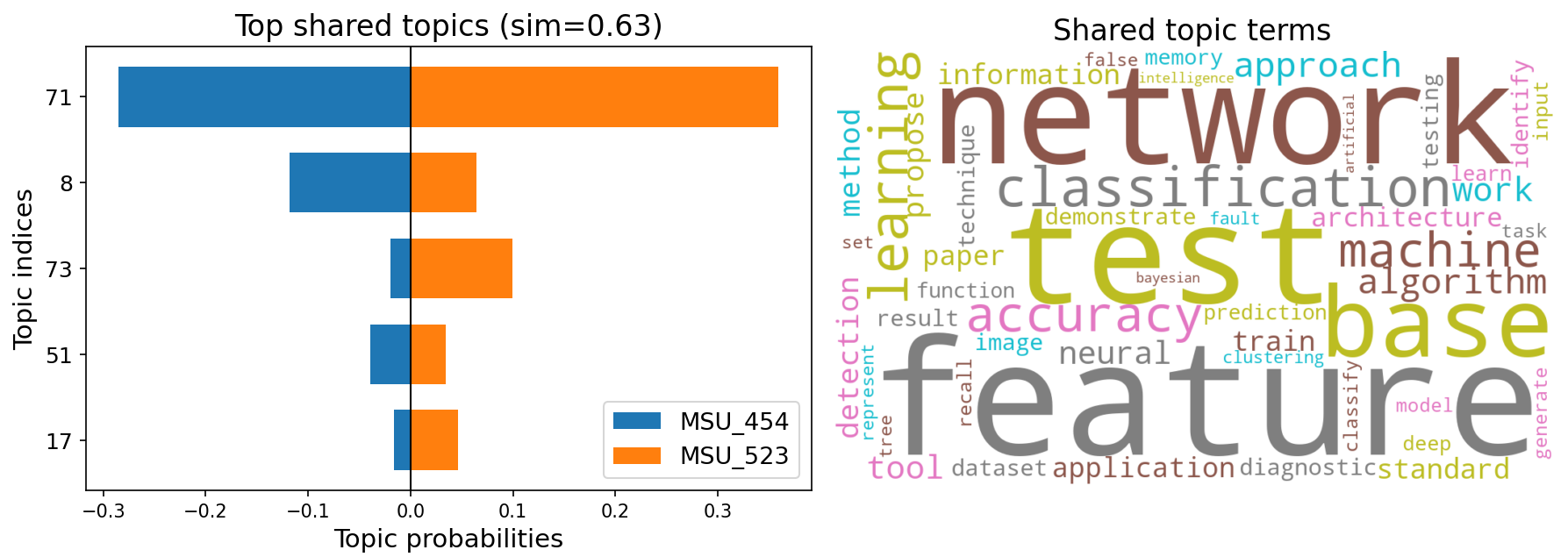}
    \caption{Topic-based explanation for a top-ranked recommendation pair}
    \label{fig:topic-expl}
\end{figure*}

We next turn to the explainability of the proposed models.
%, focusing on how recommendations can be interpreted based on their underlying representations. 
Figure~\ref{fig:topic-expl} shows an example explanation for a top-ranked recommendation pair using the topic-based approach. 
The left panel shows the top five shared topics along with the corresponding topic probabilities for each researcher and their similarity score, highlighting the primary themes driving the recommendation. The right panel shows a shared word cloud,
%constructed from these topics, 
summarizing the key overlapping terms.
%These visualizations illustrate that topic-based models are intrinsically explainable, as recommendations can be directly traced back to shared thematic structure and interpretable topic–word distributions.

We also generated a post-hoc explanation using an LLM conditioned on Faiss-based retrieval results. The same researcher pair used for Figure~\ref{fig:topic-expl} was evaluated, allowing a direct comparison between topic-level and embedding-level evidence. 
The LLM response is shown in the following box.
%The LLM response is shown in Box~\ref{box:llm_explanation}.

\begin{tcolorbox}[
    breakable,
    label={box:llm_explanation},
    title=Embedding-based Explanation Example,
    %(Uni1\_454 vs Uni1\_523),
    fontupper=\footnotesize
]

\textbf{Here are the top 3 shared research themes that suggest potential collaboration between MSU\_454 and MSU\_523.}

\vspace{0.5em}

\textbf{Theme: Hierarchical Clustering}  

\textbf{Evidence:}  
\begin{itemize}
    \item \textbf{MSU\_454:} Hierarchical fuzzy spectral clustering in social networks; hierarchical structure; sub-communities; agglomerative clustering.
    \item \textbf{MSU\_523:} MT4UML: metamorphic testing for unsupervised machine learning; metamorphic relations for K-means and agglomerative clustering algorithms.
\end{itemize}

\textbf{Explanation:}  
Both researchers focus on clustering methods that capture hierarchical structure—MSU\_454 from a network analysis perspective and MSU\_523 from a testing and validation perspective—indicating a shared interest in improving and evaluating hierarchical clustering techniques.

%\vspace{0.8em}

\textbf{Theme: Statistical Evaluation Methods}  

\textbf{Evidence:}  
\begin{itemize}
    \item \textbf{MSU\_454:} Equalized odds differences; statistical parity; exact and approximate sample-based inference.
    \item \textbf{MSU\_523:} Statistical hypothesis testing; statistical metamorphic testing for detecting faulty behavior.
    %and model correctness.
\end{itemize}

\textbf{Explanation:}  
Both researchers employ statistical techniques to assess model behavior. MSU\_454 focuses on fairness and inference metrics, while MSU\_523 applies statistical testing for validation, suggesting a shared 
methodological foundation for robust model evaluation.

%\vspace{0.8em}

\textbf{Theme: Sampling / Imbalanced Data Strategies}  

\textbf{Evidence:}  
\begin{itemize}
    \item \textbf{MSU\_454:} Protected-category oversampling; PC-SMOTE; PC-ADASYN for handling extreme imbalance.
    \item \textbf{MSU\_523:} Test input prioritization; recursive feature elimination for selecting informative samples.
\end{itemize}

\textbf{Explanation:}  
Both works address data scarcity and imbalance through sampling-based strategies, enabling more effective learning under constrained or skewed data distributions.

\end{tcolorbox}

Across both explanations, there is noticeable overlap in the high-level terminology reflected in the topic word distributions and the retrieved publication content. In particular, frequently occurring shared terms 
%such as “classification,” “machine learning,” “algorithm,” “feature,” “accuracy,” “training,” “test,” and “network” 
appear consistently across the topic-based word clouds and are also reflected in the retrieved publication evidence used for LLM generation. This indicates that both representations capture overlapping semantic signals, albeit at different levels of granularity.

Importantly, the two approaches differ in how this evidence is exposed. Topic-based explanations directly surface these terms through learned topic–word distributions, making the interpretability intrinsic to the model. In contrast, embedding-based explanations do not expose interpretable structure directly in the representation space. Instead, they rely on retrieval of relevant publications, which are then used as external evidence for the LLM to construct human-readable rationales.
% Together, these results highlight two complementary forms of interpretability: one grounded in explicit thematic structure learned by the topic model, and the other grounded in retrieval-based evidence from the embedding space.

\textbf{Limitations.}
Our study has several limitations. First, OpenAlex may not provide a complete representation of an author's research profile due to potential coverage gaps in publication metadata. Second, historical co-authorship is used as a proxy for evaluation, which may not fully reflect real-world collaboration potential, as collaboration decisions can also be influenced by social relationships, institutional factors, and other contextual barriers. Future work will explore human evaluation and additional data sources to better assess recommendation quality beyond observed collaboration links.

\section{Conclusion \& Future Work} \label{sec:conclusion}
In this work, we evaluated three families of unsupervised, content-based collaboration recommendation methods under reduced publication overlap by partially removing shared textual signals between researchers.
Our results show that TF-IDF performs best with full information but degrades substantially as lexical overlap decreases. In contrast, topic-based methods, (e.g., Clone-LDA) and embedding-based retrieval approaches remain more stable under partial information, capturing broader similarities beyond direct textual overlap.

We also compared two forms of explainability. Topic-based models provide intrinsic explanations through shared topics and term distributions, while embedding-based methods rely on retrieved publication evidence and language model-based summarization to generate human-readable justifications, with quality depending on retrieval effectiveness and model reliability.
Future work includes evaluating these methods on larger and more diverse datasets, exploring hybrid topic–embedding representations, and developing more reliable explanation methods that integrate intrinsic signals with retrieved evidence.

% \section*{Acknowledgments}
% This paper is based on work supported, in part, by NSF EPSCoR Cooperative Agreement OIA-2242802.
% Any opinions, findings, and conclusions or recommendations expressed in this material are those of the author(s) and do not necessarily reflect the views of the National Science Foundation.

% \textbf{Start of how much room you have}

% Text

% Text

% Text

% Text

% \textbf{End of how much room you have}

\balance
\bibliographystyle{IEEEtran}
\bibliography{reference}

\end{document}